\documentclass[twocolumn,preprintnumbers,prd,superscriptaddress,nofootinbib,floatfix,showpacs,showkeys]{revtex4-1}

\usepackage{amsmath}
\usepackage{amsfonts}
\usepackage{amssymb}
\usepackage{graphicx}
\usepackage{color}
\usepackage{hyperref}
\usepackage{booktabs}
\usepackage{hyperref}
\usepackage{cleveref}
\usepackage{braket}
\usepackage{mathtools}
\usepackage[rightcaption]{sidecap}
\usepackage{subfigure}
\usepackage{soul}
\usepackage{enumitem}

\definecolor{vividviolet}{rgb}{0.62, 0.0, 1.0}
\definecolor{amaranth}{rgb}{0.9, 0.17, 0.31}
\definecolor{palatinateblue}{rgb}{0.15, 0.23, 0.89}
\definecolor{brightpink}{rgb}{1.0, 0.0, 0.5}
\definecolor{cornflowerblue}{rgb}{0.39, 0.58, 0.93}
\definecolor{deepcarminepink}{rgb}{0.94, 0.19, 0.22}
\definecolor{radicalred}{rgb}{1.0, 0.21, 0.37}

\hypersetup{ linktoc=all,
    colorlinks, linkcolor={palatinateblue},
    citecolor={brightpink}, urlcolor={amaranth}
}

\newcommand{\be}{\begin{equation}}
\newcommand{\ee}{\end{equation}}
\newcommand{\bs}{\begin{split}}
\newcommand{\bea}{\begin{eqnarray}}
\newcommand{\eea}{\end{eqnarray}}

\newcommand{\bes}{\begin{subequations}}
\newcommand{\ees}{\end{subequations}}

\begin{document}

\title{Modeling black hole evaporative mass evolution via radiation from moving mirrors}

\author{Michael R. R. Good}
\email{michael.good@nu.edu.kz}
\affiliation{Department of Physics \& Energetic Cosmos Laboratory, Nazarbayev University, Kabanbay Batyr, Nur-Sultan, 010000, Kazakhstan.}

\author{Alessio Lapponi}
\email{alessio.lapponi-ssm@unina.it}
\affiliation{Scuola Superiore Meridionale (SSM), Largo San Marcellino, Napoli, 80138 Italy.}
\affiliation{Istituto Nazionale di Fisica Nucleare (INFN), Sezione di Napoli, Napoli, 80126, Italy.}

\author{Orlando Luongo}
\email{orlando.luongo@unicam.it}
\affiliation{School of Science and Technology, University of Camerino, Via Madonna delle Carceri, Camerino, 62032, Italy.}
\affiliation{Dipartimento di Matematica, Universit\`a di Pisa, Largo B. Pontecorvo, Pisa, 56127, Italy.}
\affiliation{Institute of Experimental and Theoretical Physics, Al-Farabi Kazakh National University, Almaty 050040, Kazakhstan.}
\affiliation{Istituto Nazionale di Fisica Nucleare (INFN), Sezione di Perugia, Perugia, 06123, Italy.}

\author{Stefano Mancini}
\email{stefano.mancini@unicam.it}
\affiliation{School of Science and Technology, University of Camerino, Via Madonna delle Carceri, Camerino, 62032, Italy.}
\affiliation{Istituto Nazionale di Fisica Nucleare (INFN), Sezione di Perugia, Perugia, 06123, Italy.}

\begin{abstract}
We investigate the evaporation of an uncharged and non-rotating black hole (BH) in vacuum, by taking into account the effects given by the shrinking of the horizon area. These  include the back-reaction on the metric and other smaller contributions arising from quantum fields in curved spacetime. Our approach is facilitated by the use of an analog accelerating moving mirror.  We study the consequences of this modified evaporation on the BH entropy. Insights are provided on the amount of information obtained from a BH by considering  non-equilibrium thermodynamics and the non-thermal part of Hawking radiation.
\end{abstract}

\vspace{0.5cm}


\maketitle
\tableofcontents

\section{Introduction}

Despite the impressive efforts spent on black hole (BH) thermodynamics  \cite{Hawking:1974sw, Hawking2, Beck, Beck2, Brout_1995, Ashtekar_1998, Barcel__2003, Mu_oz_de_Nova_2019}, it is still a challenge to know how a BH's mass changes in time. In the latter scenario, BH evaporation may cause a \emph{back-reaction} on the underlying spacetime metric. Consequently, several attempts to describe BH metrics encompassing back-reaction effects have been recently developed \cite{BRev1, BRev2, BRev3}, leading to no unanimous consensus on how back-reaction occurs.

Naively if a BH radiates, the horizon area \emph{shrinks} and thermal Hawking radiative power would increase. However, even the opposite perspective may be plausible, see e.g. \cite{Susskind_1992}, as quantum gravitational effects are not fully-employed \cite{BHtunnel0, Zhang_2008, ZHANG200998, BHtunnelArzano}. In addition, the modification of BH particle production is also associated with other effects, i.e., due to horizon shrinking \cite{PhysRevD.52.3512}, where, for instance, the evolution in time of the background spacetime provides a non-zero small particle count  \cite{FullingUnruhReview}.

In these scenarios, analog systems mimicking BHs, namely \emph{BH mimickers} \cite{Lemos_2008,Mimickers1,Mimickers2}, are helpful to overcome the mathematical difficulties related to time-dependent thermodynamic quantities, e.g., mass, temperature, entropy and so forth. Among all possibilities, perfectly reflecting moving mirrors in (1+1)-dimensional flat spacetime, characterized by a given trajectory, see e.g.  \cite{CW, CW2lifetime, B, Good:2016oey}, can  reproduce thermal Hawking radiation\footnote{On the other hand, semitransparent moving mirrors may exhibit quite different energy emission and particle creation \cite{DeWitt:1975ys, Davies:1976hi,Davies:1977yv,Good:2016oey, CW,good2020particle,Walker_1982,Walker:1984ya,MirrorBH4,walker1985particle,Good:2020byh,Ford:1982ct}.}.

A net advantage of mirrors consists in studying  BH radiation properties, e.g. Hawking radiation, thermodynamics, etc., without considering an underlying spacetime associated with the BH itself\footnote{Attempts towards investigating metrics considering the variation of its mass can be found in  \cite{MetricBH1,MetricBH2,MetricBH3,MetricBH4}}, as analog systems. As a consequence, by using mirrors, one can deal with BH radiation models without having a precise description of the (apparent) horizon area and/or of the BH surface gravity\footnote{For evaporating BHs, there is no Killing horizon and the concept of surface gravity is controversial. The definition of a surface gravity in these contexts is an ongoing subject of study (see e.g. \cite{Li_2021}).}.

In this work, we investigate the thermodynamic properties of a mass-varying BH adopting the BH analog provided by a thermal moving mirror. We focus in particular on the mass evolution of an evaporating BH in vacuum. The mathematical simplification of moving mirrors easily describes the mass evolution through a differential equation that can be numerically solved. We find  corrections to Hawking radiation without postulating the horizon area and/or the surface gravity. Those corrections are related to the effects that the evaporation is expected to cause to the radiation, above all, mimicking the back-reaction effect on the metric. We compare the results that we infer with those in Ref.~\cite{CW2lifetime}, where qualitative arguments for the evaporation have been discussed in view of mirrors. We debate how the expected small corrections to Hawking radiation, obtained as the BH evaporates, are of primary importance to help understand \emph{BH information loss}  \cite{BHinfpar2, BHinfpar, BHinfpar3}. Indeed, if BH radiation is not precisely thermal, then it carries some information from inside to outside the event horizon. Hence, non-thermality of BH radiation represents a landscape for the information paradox\footnote{In particular, by considering BH evaporation effects, quantum tunneling models for Hawking radiation provide a significant deviation from the thermal spectrum \cite{BHtunnel0,Zhang_2008,ZHANG200998,BHtunnelArzano}, which cause a reduction of the total BH entropy similar to the one predicted by quantum gravity \cite{Zhang_2008,BHtunnelArzano,BHentropyQG1,BHentropyQG2,BHentropyQG3}. Moreover, there exist a model-independent argument proving that the non-thermal part of Hawking radiation cannot be omitted \cite{Dvali_2013,Non-thermalitymatters1}.}. We work out the hypothesis of quasi-static processes to approximate the first thermodynamics principle by means of an effective non-equilibrium temperature. In this respect, we show that the deviations from Hawking radiation is initially small, becoming larger  as BHs evaporate. This causes a decrease of a BH's lifetime by a factor $\sim3/8$. Thus, since the effects of BH evaporation drastically affects Hawking radiation, mirrors may  confirm quantum tunneling models for Hawking radiation \cite{BHtunnel0,Zhang_2008}, showing the emitted radiation to be \emph{less entropic} than the one predicted in the literature \cite{Hawking2,Page:1976df,Page:1977um,Page2emission,Page_2005,Page_2013}. This may be interpreted by assuming part of the information can be transmitted by BH radiation. Furthermore, we emphasize in our treatment, it is possible to construct an argument for the BH age from its mass and Hawking radiation.

The paper is organized as follows. In Sec.~\ref{sec2} we explain how  moving mirror radiation emulates BHs. In Sec.~\ref{sec3} we use this analogy to study BH radiation and its mass evolution from its creation to its complete evaporation. In Sec.~\ref{sec4} we study the non-equilibrium thermodynamics of BH evaporation, adopting the quasi-static approximation. Finally, Sec.~\ref{sez5} is devoted to conclusions and perspectives of our scheme. Throughout the paper, we use Planck units $c=G=\hbar=k_B=1$.

\section{Black holes from  mirror analogy}\label{sec2}

Here we briefly review the radiation emitted by BHs and by moving mirrors. We confirm that a trajectory for a (1+1)D mirror exactly reproduces Hawking radiation emitted by a (3+1)D BH. We limit our analysis to  the emission of scalar massless particles. The discussion is split into two subsections focusing on BHs first and then the moving mirror analog.

\subsection{Black hole radiation}\label{ssec2.1}

By quantum field theory in curved spacetime, particle creation occurs whenever the background spacetime evolves in time \cite{DeWitt:1975ys}. This particle production is easy to quantify when a spacetime is flat in the infinite past and infinite future. Indeed, in this case, the normal modes of the scalar field in the infinite future (or \textit{output modes}) $\{\phi_\omega^{out}\}_\omega$ could be obtained from the ones in the infinite past (or \textit{input modes}) $\{\phi_\omega^{in}\}$ through the following Bogoliubov transformation:
\begin{equation}\label{bogtransf}
    \phi_\omega^{out}=\int_0^\infty\left(\alpha_{\omega\omega'}\phi_{\omega'}^{in}+\beta_{\omega\omega'}\phi_{\omega'}^{in\ast}\right)d\omega'\,.
\end{equation}
The non-trivial Bogoliubov coefficients $\beta_{\omega\omega'}$ are not zero, indicating that particle creation occurs from the vacuum\footnote{For relevant cosmological applications see e.g. \cite{luongo1,luongo2}.} \cite{DeWitt:1975ys,Davies:1974th}. The spectrum of particles produced is given by:
\begin{equation}\label{spectrumradiated}
    N_\omega=\int_0^\infty|\beta_{\omega\omega'}|^2d\omega'\,.
\end{equation}
Hawking calculated \cite{Hawking:1974sw} the Bogoliubov coefficients relative to a spacetime where a star collapses into a black hole. In this context, Eq.~\eqref{bogtransf} holds by considering the output modes $\{\phi_\omega^{out}\}_\omega$ as the modes outgoing from the collapsing star and the input modes $\{\phi_\omega^{in}\}_\omega$ as the modes ingoing towards it. The Bogoliubov coefficient $\beta_{\omega\omega'}$ arising from a collapsing star with mass $M$ reads:
\begin{equation}\label{BHbogcoeff}
    \beta_{\omega\omega'}=\sqrt{\frac{\omega'}{\omega}}\Gamma(1-4iM\omega)(i\omega')^{-1+4iM\omega}\,,
\end{equation}
where $\Gamma$ is the Euler gamma function. By applying the modulus square of Eq.~\eqref{BHbogcoeff} we obtain:
\begin{equation}\label{pseudoppBH}
    |\beta_{\omega\omega'}|^2=\frac{2M}{\pi\omega'}\frac{1}{e^{8\pi M\omega}-1}\,,
\end{equation}
leading to the known thermal spectrum with temperature:
\begin{equation}
    T=\frac{1}{8\pi M}.
\end{equation}
The spectrum of particles radiated, obtained from Eq.~\eqref{spectrumradiated}, is divergent because BH mass evaporation is not considered, so that the BH continues to emit forever. This is the model for what is now called an \textit{eternal BH} with exactly thermal emission. Nevertheless, by making use of wave packets, we can localize the input and output modes in a finite range of time and frequencies. In this way, Hawking proved that \cite{Hawking:1974sw}, in a finite range of time, the collapsing star emits a finite number of particles, following a thermal spectrum with temperature $1/8\pi M$. The astonishing result is that this radiation is always constant in time, with exact Planck-distributed particles originating from the collapsed star.

The fact that a BH continues to emit even when the BH is created is justified by the presence of the horizon when considering vacuum fluctuations near it \cite{Hawking:1974sw,Hawking2}.
Finally, the renormalized stress energy tensor in presence of an emitting BH was also calculated \cite{davies1976origin,Davies:1977yv}. From it, one can find the flux of energy (power) radiated by a BH as:
\begin{equation}\label{fluxHawking}
    F=\frac{1}{768\pi M^2}=\frac{\pi}{12}T^2\,.
\end{equation}
The conclusion is that, following the first quantum BH model \cite{Hawking:1974sw,Hawking2,davies1976origin,Davies:1977yv} an eternal BH emits as a $1D$ black body\footnote{To model the BH as an $n$-dimensional black body, it is sufficient to modify the pre-factor $\frac{1}{768\pi}$ from Eq.~\eqref{fluxHawking} according to the $n$-dimensional Stefan-Boltzmann constant \cite{Landsberg1989TheSC} and appropriate temperature scaling.} \cite{Bekenstein:2001tj}.\\

If we impose energy conservation, the flux of energy radiated by a BH should drain the BH mass, namely $\Dot{M}=-F$. By considering the flux \eqref{fluxHawking} we have:
\begin{equation}\label{diffeqHawking}
    \Dot{M}=-\frac{1}{768\pi M^2},
\end{equation}
providing
\begin{equation}\label{M(t)Hawking}
    M(t)=\left(M_0^3-\frac{t}{256\pi}\right)^{\frac{1}{3}},
\end{equation}
where $M_0=M(t=0)$. Following this model, the BH evaporates completely in a time
\begin{equation}\label{evatime}
    t_{ev}^H=256\pi M_0^3\,.
\end{equation}

\subsection{Mirrors and black hole analogy}\label{ssec2.2}
Another physical system providing particle production is given by an accelerating mirror. In particular, the radiation by perfectly reflecting accelerating mirrors comes from the acceleration of the boundary condition imposed by perfect reflection, providing the well-known \emph{dynamical Casimir effect} \cite{DeWitt:1975ys, Davies:1976hi,Davies:1977yv,wilson2011observation}. Let us consider a perfectly reflecting (1+1)D mirror with a generic trajectory $z(t)$. Each normal mode,  reflected back by a mirror with frequency $\omega$, i.e. $\phi_\omega^{out}$, can be written as a combination of the normal modes incoming to the mirror $\{\phi^{in}_\omega\}_\omega$ as Eq.~\eqref{bogtransf}.

The Bogoliubov coefficient $\beta_{\omega\omega'}$ is \cite{Fulling:1972md,B}
\begin{equation}
\begin{split}\label{bogcoeffmirror}
    \beta_{\omega\omega'}=\frac{1}{4\pi\sqrt{\omega\omega'}}\int_{-\infty}^{+\infty}&\exp\left(-i(\omega+\omega')t+i(\omega-\omega')z(t)\right)\\&\times\left((\omega+\omega')\Dot{z}(t)-\omega+\omega'\right)dt\,.
    \end{split}
\end{equation}

Using the renormalized stress energy tensor \cite{Davies:1977yv} one can derive the flux of energy radiated by the mirror, say to its right, as:
\begin{equation}\label{fluxgeneral}
\begin{split}
    F&=\frac{\dddot{z}(\Dot{z}^2-1)-3\Dot{z}\Ddot{z}^2}{12\pi(\Dot{z}-1)^4(\Dot{z}+1)^2}\\&=\frac{1}{12\pi}\left(\frac{\dddot{z}}{(\Dot{z}-1)^3(\Dot{z}+1)}-3\frac{\Dot{z}\Ddot{z}^2}{(\Dot{z}-1)^4(\Dot{z}+1)^2}\right)\,.
\end{split}
\end{equation}

The Carlitz-Willey trajectory corresponds to a (1+1)D trajectory and represents a simple approach to model thermal mirror trajectories. It reads
\begin{equation}\label{CarW}
    z(t)=-t-\frac{1}{\kappa}W(e^{-2\kappa t})\,,
\end{equation}
where $W$ is the Lambert function and $\kappa$ a free constant related to mirror acceleration \cite{CW}. If a mirror has this trajectory, then by Eq.~\eqref{bogcoeffmirror} we get
\begin{equation}\label{bogcoeffCWmirror}
    \beta_{\omega\omega'}=-\frac{1}{2\pi\kappa}\sqrt{\frac{\omega}{\omega'}}e^{-\frac{\pi\omega}{2\kappa}}\Gamma\left(i\frac{\omega}{\kappa}\right)\left(\frac{\omega'}{\kappa}\right)^{-i\frac{\omega}{\kappa}}\,,
\end{equation}
and its modulus square is
\begin{equation}\label{pseudoppCWmirror}
    |\beta_{\omega\omega'}|^2=\frac{1}{2\pi\kappa \omega'}\frac{1}{e^{2\pi\omega/\kappa}-1}\,.
\end{equation}
By computing from Eq.~\eqref{fluxgeneral} the flux of energy that a mirror with trajectory \eqref{CarW} radiates to its right we obtain
\begin{equation}\label{fluxCWmirror}
    F=\frac{\kappa^2}{48\pi}\,.
\end{equation}
By comparing Eq.~\eqref{pseudoppBH} with \eqref{pseudoppCWmirror}, they turn out to be equivalent, putting $\kappa=\frac{1}{4M}$. In this case, also Eqs.~\eqref{fluxHawking} and \eqref{fluxCWmirror} are the same.

Hence, a $(1+1)$-dimensional mirror with a trajectory given by Eq.~\eqref{CarW}  exactly emulates the Hawking radiation from a $(3+1)$-dimensional Schwarzschild BH with mass $\frac{1}{4\kappa}$: both in terms of particle produced and in terms of energy radiated. Considering an appropriate modification of the Carlitz-Willey trajectory \eqref{CarW}, it is possible to find an analog mirror emulating the particle production properties of a Kerr BH \cite{MirrorBH4}, a Reissman-Nordstrom BH \cite{good2020particle} and a De Sitter/AdS BH \cite{Good:2020byh}.  The exact eternal thermal emission of the Carlitz-Willey moving mirror is given by the late-time emission of the Schwarzschild mirror \cite{Good:2016oey}.

Since the modulus square of the Bogoliubov coefficients \eqref{pseudoppBH} and \eqref{pseudoppCWmirror} are the same when $M=\frac{1}{4\kappa}$, the Bogoliubov coefficients \eqref{BHbogcoeff} and \eqref{bogcoeffCWmirror} are the same up to a phase.

In the mirror framework, a phase factor on the Bogoliubov coefficient is related to a translation of the trajectory, which does not change the particle production \eqref{pseudoppCWmirror}. As a consequence, a mirror can emulate all the BH properties related to its Bogoliubov coefficients, e.g.: localized wave packets particle production \cite{B},  quantum communication properties \cite{Good:2021asq}, etc.

The Carlitz-Willey accelerated mirror has also a horizon representing the BH event horizon, i.e., from Eq.~\eqref{CarW}, we can see that the mirror approaches $z=-t$ as $t\to\infty$. This means that no particle can reach the mirror after $t=0$ and be reflected back by it. As a consequence, the information on the input particle disappears, as it does the information of a particle sent to a BH. In other words, at $t=0$, the mirror creates its horizon, as it happens for the creation of the event horizon of the BH the mirror wants to emulate.

We can notice that the energy radiated, Eq.~\eqref{fluxHawking}, does not depend upon time. Namely, the same flux arises even when the horizon is not created yet. This is due to an approximation performed in  Ref.~\cite{Hawking:1974sw}. A more realistic model should involve radiation which turns on smoothly after the creation of the horizon. The simplest of these models arises by modeling the BH as a collapsing null shell, see e.g. Ref.~\cite{massar1996} for a review.

The mirror emulating its spectrum and its energy radiated is given by the Schwarzschild mirror trajectory \cite{Good:2016oey}
\begin{equation}\label{modifiedCarW}
    z(t)=-t-\frac{1}{2\kappa}W(2e^{-2\kappa t})\,.
\end{equation}
From Ref.~\cite{Good:2016oey}, for $t<0$, the spectrum of particles and the flux of energy radiated by the mirror drops to zero exponentially as $t$ decreases. On the contrary, for $t>0$, both those quantities go exponentially to the Hawking ones, Eqs.~\eqref{BHbogcoeff} and \eqref{fluxHawking} respectively, as $t$ increases. The deviation with respect to the particle spectrum and energy radiated, Eq.~\eqref{fluxHawking}, predicted by Hawking drops as $\sim e^{-t/M}$. Hence, the radiation could be considered as completely `turned on' when $t\gtrsim M$.

In the next section, for simplicity, we consider a BH starting to evaporate only when $t\gtrsim M$, or fully turned on. We see that, if the initial mass of the BH is large enough, the \textit{turning on period} occurs in a time negligible with respect to the evaporation period, justifying why this approximation may hold.

\section{Black hole evaporation}\label{sec3}

In the following, we generalize the Carlitz-Willey trajectory of Eq. \eqref{CarW} by taking $\kappa$ time dependent. Thus, from the BH mirror analogy, $\kappa=\frac{1}{4M}$,  a variation of $\kappa$ induces a variation over the BH mass and also represents a class of trajectories quite different from the standard Carlitz-Willey.

By imposing energy conservation, we can thus find $M(t)$, with the corresponding flux deviating from Eq.~\eqref{fluxHawking} as due to the time dependence of $\kappa$.

This deviation can be easily related to the effects expected to slightly modify Hawking radiation during BH evaporation, such as the back-reaction on the metric or the shrinking of the horizon modifying local boundary conditions. In all these situations, we expect departures from genuine equilibrium thermodynamics, in favor of non-equilibrium effects that we will discuss later in the text.

\subsection{Modeling BH evaporation with mirrors}

As discussed at the end of Sec.~\eqref{ssec2.2}, the black hole radiation turns on smoothly once the black hole is created. For simplicity, we consider that the black hole starts to evaporate once the radiation is fully turned on. In this way, we can consider the analog mirror to follow a generalization of the standard Carlitz-Willey trajectory,  Eq.~\eqref{CarW}, namely
\begin{equation}\label{generalizedCW}
    z(t)=-t-4 M(t)W\left(e^{-\frac{t}{2M(t)}}\right)\,.
\end{equation}
The time $t=0$ corresponds to the time at which the BH starts to evaporate - this makes $t\sim -4M_0$ the time at which the BH has been created (see the discussion at the end of Sec.~\ref{ssec2.2}).

We define $M_0=M(t=0)$ as the initial mass hold by the underlying BH. To generalize the flux, we should plug Eq.~\eqref{generalizedCW} and its time derivatives into Eq.~\eqref{fluxgeneral}. In this way, we can obtain a general expression for the flux\footnote{The expression is not explicitly reported because it is too cumbersome.} $F=F(t,M,\dot{M},\Ddot{M},\dddot{M})$. In so doing, from the energy conservation $\Dot{M}=-F$, we get a third order ordinary differential equation:
\begin{equation}
    \dot{M}=-F(t,M,\dot{M},\Ddot{M},\dddot{M})\,,
\end{equation}
giving the evolution of the mass $M(t)$ from $t=0$ to the complete evaporation time $t_{ev}$.

To simplify the flux expression $F(t,M,\dot{M},\Ddot{M},\dddot{M})$, we employ two ranges of time:
\begin{itemize}
    \item[-] $0\le t \le t_0$, where we consider negligible the deviations from the Hawking flux \eqref{fluxHawking}.
    \item[-] $t_0<t<t_{ev}$, where the corrections to the Hawking flux \eqref{fluxHawking} given by the BH evaporation becomes non-negligible.
\end{itemize}
The time $t_0$ is fixed as $t_0\gg 2M_0$, implying $t\gg 2M(t)$ for $t_0<t<t_{ev}$, since $M(t)$ decreases in time as the BH evaporates.

Hence, in the generalized Carlitz-Willey trajectory, we may approximate $W(\exp(-t/M(t)))\sim \exp(-t/M(t))$. Thus, for $t\le t_0$ we easily recover the Hawking flux of Eq. \eqref{fluxHawking}, whereas for $t>t_0$ the flux can be computed from Eq.~\eqref{fluxgeneral}, applying the approximation $t\gg 2M(t)$. Hence, the flux of energy radiated from $t=0$ to $t=t_{ev}$ is:
\small
\begin{equation}\label{fluxtoogeneral}
    \begin{cases}
    F=\frac{1}{768\pi M^2}\hspace{1 cm}\text{for}\; 0<t\le t_0\,;\\
    \begin{split}
    F=&\frac{1}{192\pi\left(1-\frac{\Dot{M}}{M}t\right)^2}\left(3\frac{\Ddot{M}^2}{M^2}t^2+\frac{1}{4M^2}\left(1-\frac{\Dot{M}}{M}t\right)^4\right.\\&\left.+2\frac{\dddot{M}}{M}t\left(1-\frac{\Dot{M}}{M}t\right)-12\frac{\Ddot{M}\Dot{M}}{M^2}t\right)\hspace{0.5 cm}\text{for}\;t>t_0\,.
    \end{split}
    \end{cases}
\end{equation}\normalsize
To apply this simplification, we must ensure $\dot{M}$ to be  negligible for $t\le t_0$, i.e., $-\dot{M}(t_0)\ll M(t_0)$.

Since, for $t\le t_0$, Eq.~\eqref{M(t)Hawking} is valid, then at $t=t_0$ we obtain
\begin{equation}
    |\dot{M}(t_0)|=\frac{1}{768\pi M^2(t_0)}\ll M(t_0)\,.
\end{equation}
From Eq.~\eqref{M(t)Hawking}, that is valid for $t\le t_0$, we can therefore evaluate $M(t_0)$. Thus, the condition $-\dot{M}(t_0)\ll M(t_0)$ becomes
\begin{equation}\label{approxt0}
    t_0\ll 256\pi M_0^3-\frac{1}{3}.
\end{equation}
Summing up, we need to choose a time, $t_0$, such that $2M_0\ll t_0\ll 256\pi M_0^3-\frac{1}{3}$. So, in order to have a $t_0$ satisfying this condition, we need an initial mass, $M_0$, large enough\footnote{For instance, $M_0\sim5$ ensures the existence of a $t_0$ which is $100$ times smaller than $256\pi M_0^3$ and $100$ times larger than $2M_0$, making the approximation valid.}.

\subsection{Evaluating the mass evolution}

Afterwards, to evaluate the function $M(t)$ we impose the energy condition, $\dot{M}(t)=-F$.

In this way, Eq.~\eqref{fluxtoogeneral} becomes a third order, non linear differential equation
\begin{equation}\label{massevotoogeneral}
    \begin{cases}
    \Dot{M}=\frac{1}{768\pi M^2}\hspace{1 cm}\text{for}\; 0<t<t_0\,.\\
    \begin{split}
    \Dot{M}=&-\frac{1}{192\pi\left(1-\frac{\Dot{M}}{M}t\right)^2}\left(3\frac{\Ddot{M}^2}{M^2}t^2+\frac{1}{4M^2}\left(1-\frac{\Dot{M}}{M}t\right)^4\right.\\&\left.+2\frac{\dddot{M}}{M}t\left(1-\frac{\Dot{M}}{M}t\right)-12\frac{\Ddot{M}\Dot{M}}{M^2}t\right)\hspace{0.5 cm}\text{for}\;t>t_0\,.
    \end{split}\\
    M(t=0)=M_0\,.
    \end{cases}
\end{equation}

\begin{figure}
\centering
\includegraphics[scale=0.65]{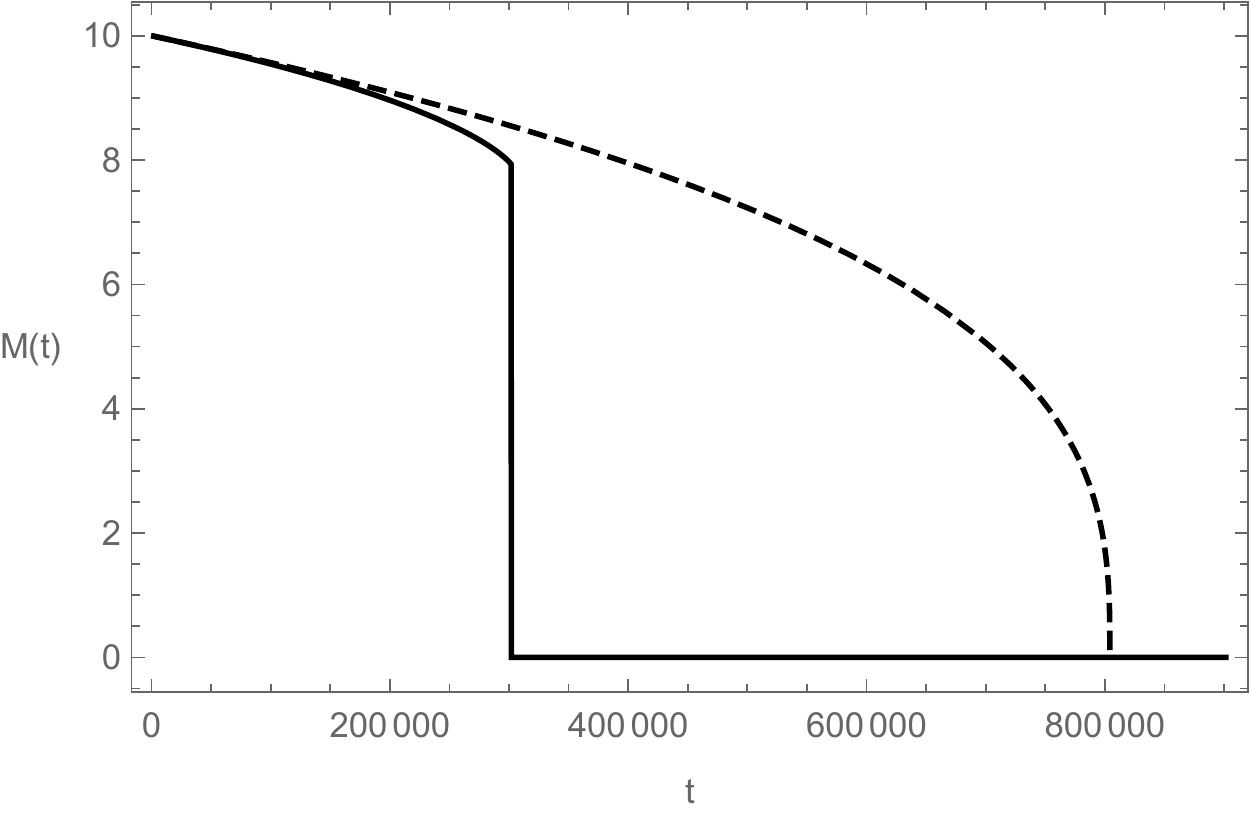}
\caption{Plot of the numerical solution of $M(t)$ from Eq.~\eqref{massevotoogeneral} (continuous line) and Hawking solution for $M(t)$ following Eq.~\eqref{M(t)Hawking} (dashed line) considering $M_0=10$ and $t_0=200 M_0$.}
\label{MassWithStiffness}
\end{figure}

The numerical solution of Eq.~\eqref{massevotoogeneral} is drawn in Fig.~\ref{MassWithStiffness}, where $t_0=200 M_0$ and $M_0=10$ were considered, having $t_0\sim 2000$. The period of time $0<t<t_0$, in which the evaporation effects on the radiation are neglected, is very small with respect to the overall evaporation period, i.e., one part over a thousand. This is what we wanted, since we want to study the deviations from the Hawking radiation \eqref{fluxHawking} in a period of time as large as possible.

In Sec.~\ref{ssec2.2}, the ‘‘turning on period", namely the period in which the radiation turns on after BH creation, is $\sim 4M_0$, being very small than periods of time here-considered. Consequently, we can ignore the turning on period, identifying the time $t=0$ as the time at which the horizon of the BH originates as consequence of star collapse.

From the numerical solution in Fig. \ref{MassWithStiffness}, a sudden drop of the mass occurs at $t=t_c$, namely at a  \textit{critical time}, unavoidable for any initial mass value. To analytically explain this sharp behaviour, we now approximate Eq.~\eqref{massevotoogeneral} at the range of times $t_0<t<t_c$.

To do so, we first estimate the magnitude of $\Dot{M}$, $\Ddot{M}$ and $\dddot{M}$ when there are no evaporation effects on the radiation. Using Eq.~\eqref{diffeqHawking}, we get:
\begin{equation}\label{massderivativesorder}
    \begin{cases}
    &\dot{M}\sim10^{-3}/M_0^2\,;\\
    &\Ddot{M}\sim10^{-6}/M_0^5\,;\\
    &\dddot{M}\sim10^{-9}/M_0^8\,.
    \end{cases}
\end{equation}
Considering the evaporation effects on radiation, the derivatives of the mass \eqref{massderivativesorder} are expected to increase in magnitude. However, from Fig.~\ref{MassWithStiffness}, we see that such increase is relatively small. Bearing this in mind,  we study the orders of the terms at the r.h.s. of Eq.~\eqref{massevotoogeneral}, as  $t>t_0$, using Eqs.~\eqref{massderivativesorder}.
\begin{itemize}
    \item[-] The term proportional to $\frac{\Ddot{M}^2t^2}{M^2}$ has order $\sim 10^{-8}/M_0^6$. The denominator $(1-\frac{\Dot{M}}{M}t)^2$ decreases the magnitude of this term as $t$ increases\footnote{Consider that $\Dot{M}$ is forced to be negative, so $(1-\frac{\Dot{M}}{M}t)$ is always larger than $1$, increasing as  time increases.}. The same thing is valid for the third and last term, respctively.
    \item[-] The second term is $\sim1/M_0^2$ and  simplifies as $\frac{1}{M^2}(1-\frac{\Dot{M}}{M}t)^2$ with the denominator, increasing \emph{de facto} with time.
    \item[-] The third term has order $10^{-7}/M^7$, with magnitude decreasing as  time increases.
    \item[-] The last term has order $10^{-7}/M^6$, with magnitude decreasing as  time increases.
\end{itemize}

\begin{figure}
    \centering
    \includegraphics[scale=0.6]{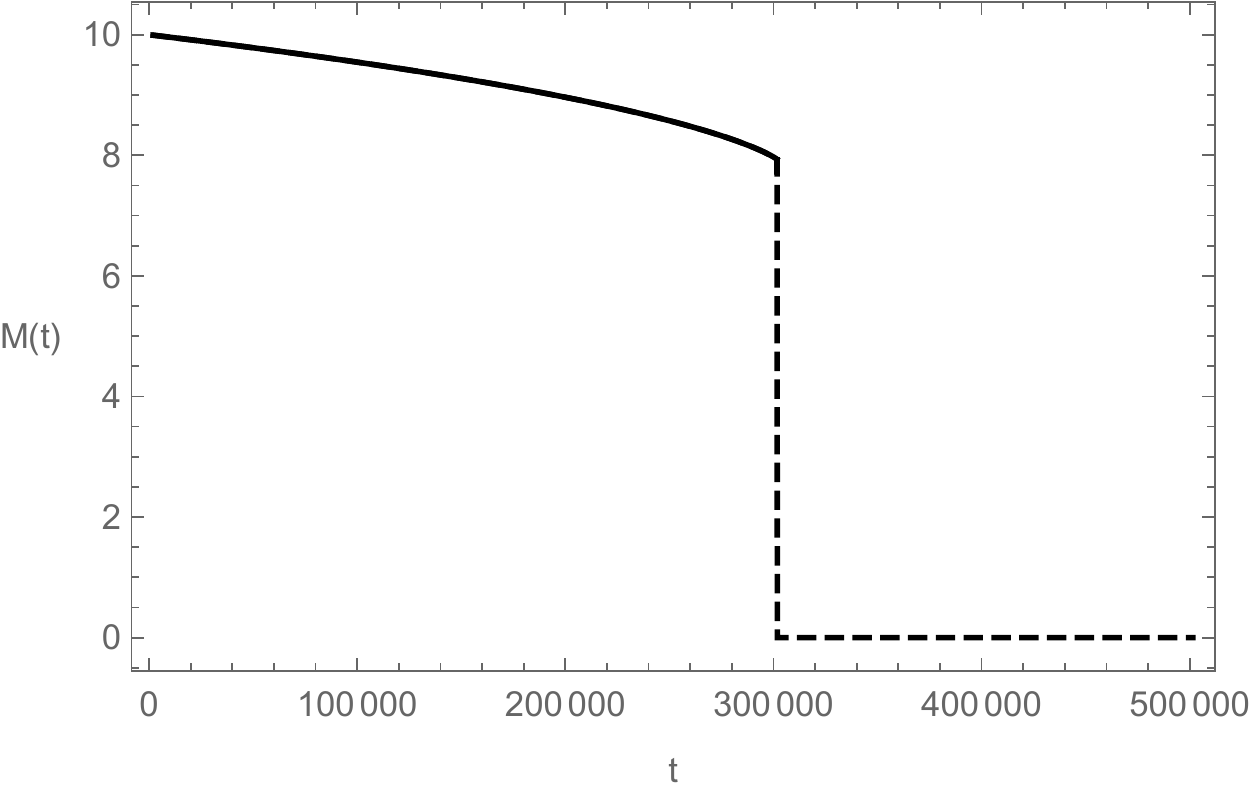}
    \caption{Comparison between the solution of the approximated Eq.~\eqref{diffeqimplapprox} (thick line) and the solution of Eq.~\eqref{massevotoogeneral} (dashed line).}
    \label{ComparisonCurves}
\end{figure}

From this analysis, since $M_0$ cannot be small, we  conclude that the second term, i.e., $\frac{1}{4M}\left(1-\frac{\Dot{M}}{M}t\right)^2$ is dominant before $t_c$, leading to an approximation of Eq.~\eqref{massevotoogeneral} of the kind:
\begin{equation}\label{diffeqimplapprox}
\Dot{M}=-\frac{1}{768\pi M^2}\left(1-\frac{\Dot{M}}{M}t\right)^2.
\end{equation}
The validity of Eq.~\eqref{diffeqimplapprox} before $t_c$ is confirmed in Fig.~\ref{ComparisonCurves}, where   the solutions of Eqs.~\eqref{massevotoogeneral} and \eqref{diffeqimplapprox} effectively coincide before $t_c$.

\subsection{Interpreting the critical time}

The issue of inferring the physical consequences of the above-defined critical time is challenging but it helps to justify the existence of critical time via the use of the simplified differential Eq.~\eqref{diffeqimplapprox}.  Indeed, Eq.~\eqref{diffeqimplapprox} can be written as
\begin{equation}\label{diffeqapprox}
    \Dot{M}=-\frac{384\pi M^4}{t^2}+\frac{M}{t}+\frac{384\pi M^4}{t^2}\sqrt{1-\frac{t}{192\pi M^3}}\,,
\end{equation}
whose solution exists if $t\le192\pi M^3(t)$. So, as $t\rightarrow192\pi M^3$, $\ddot M(t)$ diverges as $\sim \left(1-\frac{t}{192\pi M^3}\right)^{-1/2}$. As a consequence, also the third time derivative of the mass diverges. So, in a neighborhood of $t=192\pi M^3(t)$, the first, third and fourth terms at the right hand side of the second of Eq.~\eqref{massevotoogeneral} suddenly increase, becoming dominant and making $M(t)$ to drop sharply as Fig.~\eqref{MassWithStiffness} shows.

At this point, we can associate $t_c$ to the time at which the square root of Eq.~\eqref{diffeqapprox} nullifies.
Further, we define the \textit{critical mass} as $M_c\coloneqq M(t_c)$, having this relation:
\begin{equation}\label{relationCritValues}
    t_c=192 \pi M_c^3;
\end{equation}
\begin{figure}
    \centering
    \includegraphics[scale=0.6]{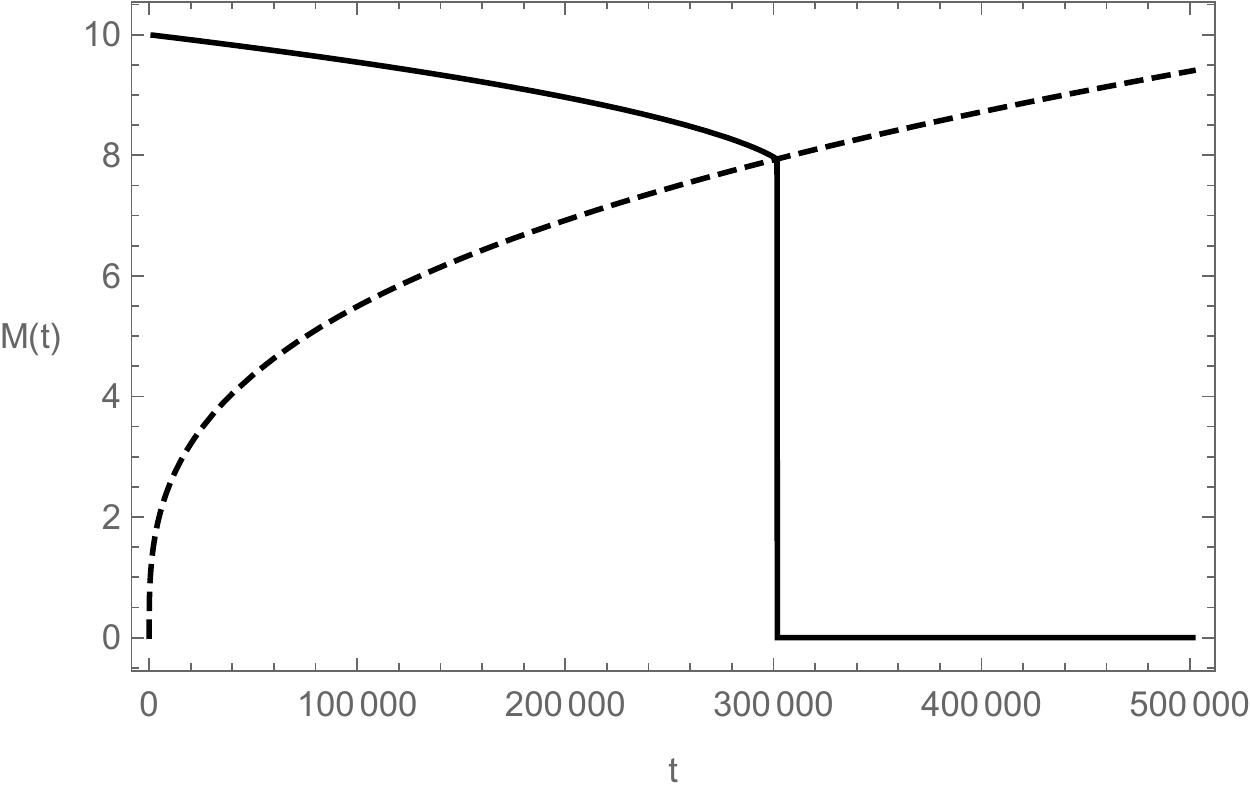}
    \caption{This figure shows that the critical point of the solution of Eq.~\eqref{massevotoogeneral} for $M_0=10$ (thick line) satisfies the relation \eqref{relationCritValues}, since the critical point $(t_c,M_c)$ lies on the curve $t=192\pi M^3$ (dashed line).}
    \label{CritCross}
\end{figure}
Fig.~\ref{CritCross} shows that the critical point lies on the curve $t=192\pi M^3$ confirming the relation \eqref{relationCritValues} between critical time $t_c$ and critical mass $M_c$.  Numerically, taking a sample of initial masses stepping by $\Delta M_0=1$ from $M_0=10$ to $M_0=50$, the critical mass becomes $M_c\sim0.7937 M_0\simeq 2^{-1/3}M_0$.

With this information, we can give a value to the critical time and mass:
\begin{equation}\label{empcritvalues}
    M_c\simeq\frac{M_0}{\sqrt[3]{2}} \Longrightarrow t_c\simeq96\pi M_0^3=\frac{3}{8}t_{ev}^H,
\end{equation}
where $t^H_{ev}$ is the evaporation time predicted by Hawking \eqref{evatime}.

After the critical time $t_c$, the mass drops to zero in a finite but very short time, that we can neglect. Thus, the new evaporation time of the BH is modified as $t_{ev}\sim \frac{3}{8}t_{ev}^H$, demonstrating the BH evaporates faster than the standard Hawking case when accounting for the mass evaporation effects on the radiation. This fact reduces the evaporation time by a factor $\sim 3/8$.

The mass behavior after $t_c$ requires a physical interpretation.

\begin{itemize}
    \item[-] For instance, a possible justification may include quantum gravity effects. Indeed, since $\Dot{M}$ increases sharply after the critical time, $|\Dot{M}|$ quickly reaches order $M$ and likely, as this fact occurs, quantum gravity effects cannot be neglected. Hence, our modeling predicts that, when considering the mass evaporation of a black hole in vacuum and the effects the evaporation induces to the metric, quantum gravity effects have to be considered when the black hole mass reaches $2^{-1/3}$ of its initial value.
    \item[-] Analytically, the sharp mass drop is due to the sudden increasing of $-\Ddot{M}$, i.e. of the \emph{mass loss acceleration}, analyzed after Eq.~\eqref{diffeqapprox}. It appears evident that the accelerated behavior resembles a jet-like form, similar to Fermi processes \cite{Luongo:2021pjs}, where the mass loss acceleration does not smoothly behave, leading to uncontrolled astrophysical processes. This interpretation may be framed in more practical physical scenarios related to compact object, leaving open the possibility to model those objects by accelerated mirrors.
\end{itemize}

We give these explanations for the sake of completeness, however they lie beyond the main purposes of our work, however interesting they may be for future investigations.

\subsection{Mass evolution at early times}
In this subsection, we provide an approximation for the differential Eq. \eqref{diffeqapprox} valid when $t\gtrsim t_0$. In this way, we obtain an analytic expression for $M(t)$, providing an explanation for how the mass evaporation of the black hole modifies the Hawking radiation at early times.

From the condition \eqref{approxt0}, by considering $t\gtrsim t_0$ we expect also $t\ll 192\pi M^3(t)$. Hence, the square root in the right hand side of Eq.~\eqref{diffeqapprox} can be expanded up to third order as:
\begin{equation}\label{squarerootexpanded}
\begin{split}
    \sqrt{1-\frac{t}{192\pi M^3}}\sim \,&1-\frac{t}{384\pi M^3}-\frac{2t^2}{9\cdot(256\pi)^2 M^6}\\&-\frac{4t^3}{27\cdot(256\pi)^3 M^9}+\ldots\,.
\end{split}
\end{equation}
Considering the first two terms of the expansion \eqref{squarerootexpanded}, Eq.~\eqref{diffeqapprox} becomes $\Dot{M}=0$ i.e. the case in which there is no evaporation. Considering the first three terms of \eqref{squarerootexpanded} we get exactly the known differential equation \eqref{diffeqHawking} with the known solution \eqref{M(t)Hawking}. To provide a first correction on $M(t)$ given by the evaporation effects, we can consider also the fourth term of the expansion \eqref{squarerootexpanded}. In this case Eq.~\eqref{diffeqapprox} becomes:
\begin{equation}\label{diffeqstepfurther}
    \Dot{M}=-\frac{1}{768\pi M^2}-\frac{2t}{9\cdot (256\pi)^2 M^5}.
\end{equation}
This can be rewritten in terms of $t(M)$:
\begin{equation}\label{diffeqstepfurthert(M)}
    \frac{dt}{dM}=-\frac{768\pi M^2}{\left(1+\frac{t}{384\pi M^3}\right)}\,.
\end{equation}
By considering $\left(1+\frac{t}{384\pi M^3}\right)^{-1}\sim1$ then the solution \eqref{M(t)Hawking} is restored. Hence, to study its first order deviation, we expand the latter to first order for $t\ll 256\pi M^3$, namely $\left(1+\frac{t}{384\pi M^3}\right)^{-1}\sim1-\frac{t}{384\pi M^3}$. In this way, Eq.~\eqref{diffeqstepfurthert(M)} becomes the linear differential equation:
\begin{equation}\label{diffeqtosolve}
    \frac{dt}{dM}-\frac{2t}{M}=-768\pi M^2\,.
\end{equation}
From Eq.~\eqref{massevotoogeneral}, for $0<t\le t_0$, the mass evolution is given by Eq.~\eqref{M(t)Hawking}. Since Eq.~\eqref{diffeqtosolve} is valid for $t\gtrsim t_0$, the initial condition for it is defined using Eq.~\eqref{M(t)Hawking} at $t_0$, i.e.:
\begin{equation}\label{incondtosolve}
    t\left(\left(M_0^3-\frac{t_0}{256\pi}\right)^{1/3}\right)=t_0\,.
\end{equation}
The solution of Eq.~\eqref{diffeqtosolve} with the condition \eqref{incondtosolve} is:
\begin{equation}\label{generalSol}
\begin{split}
    t(M)=&\frac{1}{M^2}\left(\frac{2}{5}t_0+\frac{768\pi M_0^3}{5}\right)\left(M_0^3-\frac{t_0}{256\pi}\right)^{2/3}+\frac{768\pi M^3}{5}.
    \end{split}
\end{equation}
Here, $t_0$ is arbitrary, but since $t_0\ll t_{ev}^H$ we expand the two factors in Eq.~\eqref{generalSol} depending on $t_0$:
\begin{equation*}
    \begin{split}
    &\left(\frac{2}{5}t_0+\frac{768\pi M_0^3}{5}\right)\left(M_0^3-\frac{t_0}{256\pi}\right)^{2/3}\\&=\frac{768\pi M_0^5}{5}\left(1+\frac{t_0}{384\pi M_0^3}\right)\left(1-\frac{t_0}{256\pi M_0^3}\right)^{2/3}\\&
    \sim \frac{768\pi M_0^5}{5}\left(1+\mathcal{O}\left(\left(\frac{t_0}{t_{ev}^H}\right)^2\right)\right)\,,
    \end{split}
\end{equation*}
proving that $t_0\ne0$ provides a second order deviation from the Hawking case. However, since we have considered only first deviations from the Hawking mass evolution \eqref{M(t)Hawking}, we can neglect the contribution of $t_0\ne0$.\\
In this way, the solution of \eqref{diffeqtosolve} becomes:
\begin{equation}\label{firstcorrectedsolution}
    t(M)=\frac{768\pi}{5}\left(\frac{M_0^5-M^5}{M^2}\right).
\end{equation}
In contrast, the one without evaporation effects \eqref{M(t)Hawking} reads:
\begin{equation}
  t(M)=256\pi(M_0^3-M^3),
\end{equation}
Summarizing, Eq.~\eqref{firstcorrectedsolution} expresses the correct behavior of the mass in time when $t\ll t_{ev}$ i.e. when evaporation effects are small but different than $0$. One can study further corrections of Eq.~\eqref{M(t)Hawking} by considering higher orders of the expansion \eqref{squarerootexpanded}.

\subsection{Dynamical behavior of mirrors at intermediate stages}

By virtue of the general mass loss behavior,  Eq.~\eqref{massevotoogeneral}, one can infer how the mirror trajectory evolves throughout the evolution of our dynamical system.

\begin{figure}
    \centering
    \includegraphics[scale=0.7]{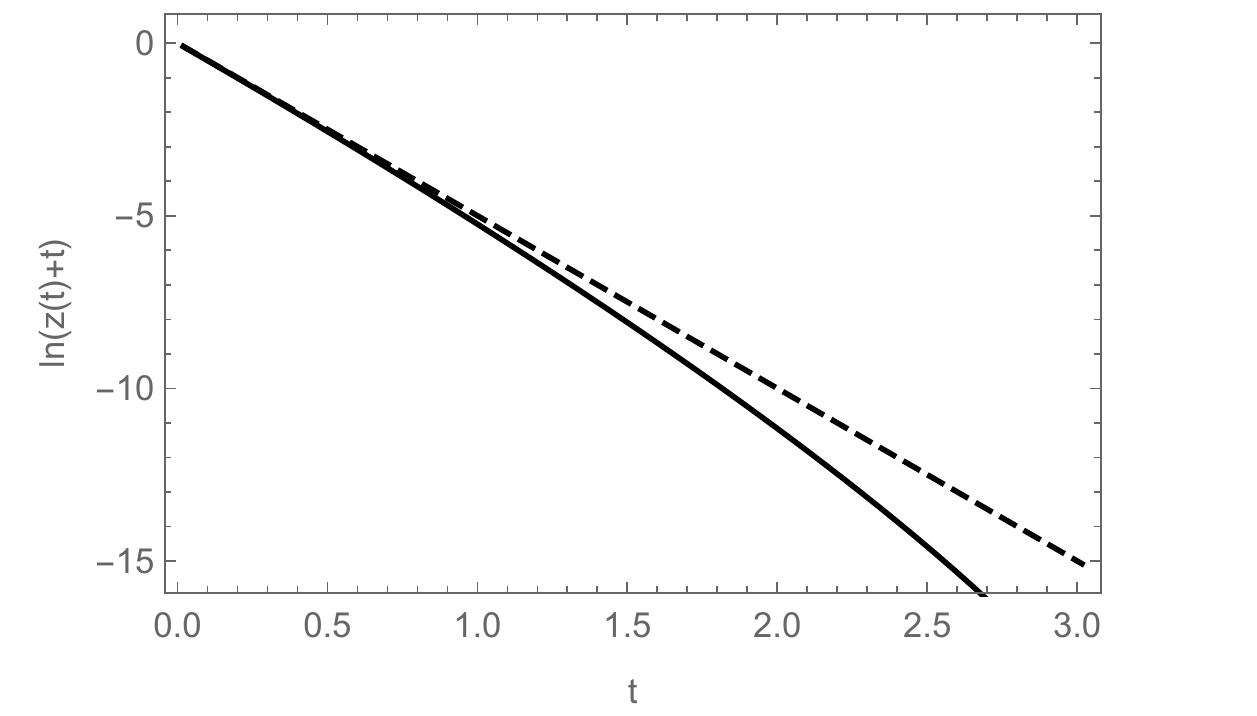}
    \caption{Trajectory \eqref{generalizedCW} of the mirror emulating an evaporating BH (thick line) and trajectory \eqref{CarW} emulating an eternal BH (dashed line). The numbers on the axes $x$ and $y$ are in powers of $10^5$ and $10^3$, respectively}
    \label{trajcomparisons}
\end{figure}

For our purposes, the trajectory of the mirror \eqref{generalizedCW} has been defined only in the restricted range of times $0<t<t_{ev}$
as a modification of the Carlitz-Willey trajectory \cite{CW},  Eq.~\eqref{generalizedCW}, with $M$ time-dependent and following the differential Eq.~\eqref{massevotoogeneral}. The modification of the Carlitz-Willey trajectory is shown in Fig.~\ref{trajcomparisons}. In our trajectory, the mirror approaches the asymptote $z=-t$ faster than the normal Carlitz-Willey trajectory. However, the difference between the two is vanishingly small, namely, of an order $e^{-2t/M_0}-e^{-2t/M(t)}$.
Now, we can argue which consequences occur to the mirror at the critical time $t_c$. To do so, we know that, near $t_c$, $\Ddot{M}$ suddenly increases proportionally to $\left(1-t/t_c\right)^{-1/2}$. Thus, using Eq.~\eqref{generalizedCW} and $ t\gg2M(t)$, we can compute the mirror velocity and  acceleration with respect to an external observer, respectively, as
\begin{equation}\label{appr1der}
    \Dot{z}(t)\sim -1+2\left(1-\frac{\Dot{M}}{M}t\right)e^{-\frac{t}{2M}};
\end{equation}
\begin{equation}\label{appr2der}
    \Ddot{z}(t)\sim -2\left(\frac{\Ddot{M}}{M}t+\frac{1}{2M}\left(1-\frac{\dot{M}}{M}t\right)^2\right)e^{-\frac{t}{2M}}\,.
\end{equation}
Looking at Eq.~\eqref{appr2der}, the acceleration drops always as an exponential $e^{-2\kappa t}$. As $t$ approaches $t_c$, the first term of Eq.~\eqref{appr2der} dominates and the acceleration of the mirror becomes proportional to:
\begin{equation}
    \Ddot{z}(t)\propto \frac{e^{-\frac{M}{2t}}}{\sqrt{1-\frac{t}{t_c}}}\,.
\end{equation}
For $t\lesssim t_c$, the acceleration of the mirror is vanishingly small. However, $e^{-t/2M}$ never goes precisely to $0$, but the square root in the denominator does. As a consequence, when $t_c-t$ is really close to zero, the acceleration, from being vanishingly small, suddenly diverges. As this happens, the mirror reaches its asymptote $z=-t$ suddenly. In particular, this occurs when the BH completely evaporates, namely for $M\to0$) - as it can be easily verified from Eq.~\eqref{CarW}. This is consistent with the fact that, at the moment of the evaporation $t_{ev}$, both the BH horizon and the mirror horizon disappear (see the discussion in Sec.~\ref{ssec2.2} on the horizon analogy).

Once the BH has evaporated, the mirror should be static in order to reproduce a flat spacetime where signals are not redshifted \cite{Walker_1982, CW2lifetime, Good_2018}. This implies that $\Dot{z}(t_{ev})$ should be zero, but this means that we can assert the behaviour of $\Dot{M}$ when $M$ approaches zero. In fact, taking the velocity of the mirror as Eq.~\eqref{appr1der} and imposing it to be null at $t=t_{ev}$, we obtain:
\begin{equation}
    \Dot{M}=-\frac{M}{t_{ev}}\left(\frac{e^{\frac{t}{2M}}}{2}-1\right)\,.
\end{equation}
In conclusion, in the limit $M\to 0$, $\Dot{M}$ diverges to $-\infty$ asymptotically to:
\begin{equation}
    \Dot{M}\sim -\frac{M}{t_{ev}}e^{\frac{t_{ev}}{2M}}\,.
\end{equation}

Comparing the behaviour of our mirror w.r.t. to Ref.~\cite{CW2lifetime}, the deceleration of our mirror to become static is very fast. As a future perspective, one can try to use the formalism of Ref.~\cite{CW2lifetime} to slow down the mirror deceleration after $t_c$.

\section{Mirror thermodynamics and  black hole analogy}\label{sec4}

In this section, we would like to study the thermodynamics of the evaporating BH modeled in Sec.~\ref{sec3}. We aim to know if the entropy released by the BH during its evaporation is less than the one predicted by Bekenstein and Hawking \cite{Hawking:1974sw,Beck}.

Eternal BH Hawking radiation has a thermal spectrum, and the radiation of the BH modeled in Sec.~\ref{sec3} deviates from the thermal one. We expect the non-thermal part of the radiation to contain some information, unavailable otherwise, about the BH. It is worth pointing out that, with the mirror model, we cannot find exact results for the thermality of the spectrum of particles radiated\footnote{Indeed, the Bogoliubov coefficient $\beta_{\omega\omega'}$ (Eq.~\eqref{bogcoeffmirror}) gives the spectrum of the overall particles radiated during the evaporation, without time-dependence. Moreover, Ref.~\cite{B} shows that, in the mirror context, the study of time-dependent particle production through localized wave packets may give controversies.}. However, the expressions obtained in the previous section allow a physically reliable assumption for the non-thermal part of the flux radiated. The price to be paid is that the final results are dependent on an unknown index. Nevertheless, we are able to restrict this parameter to a small range by putting ourselves in a quasi-static regime.

\subsection{Non-thermality}

From Sec.~\ref{sec2}, the flux of energy radiated by an evaporating black hole, i.e. its power, is given by Eq.~\eqref{fluxtoogeneral}. When $0\le t\le t_0$, the power radiated is exactly the one predicted by Hawking, Eq. \eqref{fluxHawking}. For this reason, we can suppose the radiation to be completely thermal in this range of times. When $t_0<t<t_{ev}$, the radiation deviates from the Hawking one \eqref{fluxHawking}, by
\begin{equation}\label{hawkingdeviation}
    \Delta F=F-\frac{1}{768\pi M^2}\,.
\end{equation}
By Eq.~\eqref{fluxtoogeneral} we have in particular\small
\begin{equation}\label{nonthermalflux}
    \begin{cases}
    \Delta F=0\hspace{0.5 cm}\text{for}\; 0<t\le t_0\,;\\
    \begin{split}
    &\Delta F=\frac{1}{192\pi\left(1-\frac{\Dot{M}}{M}t\right)^2}\left(3\frac{\Ddot{M}^2}{M^2}t^2+2\frac{\dddot{M}}{M}t\left(1-\frac{\Dot{M}}{M}t\right)\right.\\&\left.-12\frac{\Ddot{M}\Dot{M}}{M^2}t\right)+\frac{1}{768\pi M^2}\left(\frac{\Dot{M}^2}{M^2}t^2-2\frac{\Dot{M}}{M}t\right)\hspace{0.5 cm}\text{for}\;t>t_0\,.
    \end{split}
    \end{cases}
\end{equation}\normalsize
We notice that $\Delta F$ nullifies whenever $\dot{M}=0$,  whereas when $\Delta F=0$ the spectrum turns out to be exactly thermal. So, the non-thermality of Hawking radiation is expected to be proportional to $\Delta F$. In particular we state that the BH power is composed of a thermal contribution $F_{\textrm{th}}$ and a non-thermal one $F_{\textrm{no-th}}$. We thus have
\begin{equation}\label{ans1}
    F=F_{\textrm{th}}+F_{\textrm{no-th}}\,.
\end{equation}
We know that the Hawking flux $1/(768\pi M^2)$ gives an exact thermal contribute, so that this term is included in $F_{th}$. However, the possibility that part of $\Delta F$ gives a small thermal contribute cannot be excluded\footnote{We can imagine the overall spectrum of particles radiated as the superposition between the thermal one given by $1/(768\pi M^2)$ and an unknown contribute given by $\Delta F$. However, the unknown contribute can modify the thermal spectrum created by $1/(768\pi M^2)$ such that another thermal spectrum, with different temperature, arises.}. Since the non-thermality of the spectrum must be proportional to the deviation from Hawking radiation $\Delta F$, we write:
\begin{equation}\label{ans2}
    F_{\textrm{no-th}}=\alpha \Delta F\,,
\end{equation}
where $\alpha\in[0,1]$ is an unknown parameter that acts to quantify how much the deviations from Hawking power are non-thermal.
The thermal part of the radiation (the one giving the thermal spectrum) is then:
\begin{equation}
    F_{\textrm{th}}=\frac{1}{768\pi M^2}+(1-\alpha)\Delta F\,.
\end{equation}
Thus, without considering the Bogoliubov coefficients  to study the spectra of particles radiated~\cite{B}, the BH thermodynamics is then studied considering the above  parameter $\alpha$, restricted to be close to unity in order to fulfill the quasi-static regime, as we clarify in the subsection below.

\subsection{Temperature and entropy of an evaporating BH}

Consider the first law of thermodynamics,
$dE=dQ-d\Theta$, where $\Theta$ is a generic quantity associated with a loss of energy and not given by heat exchange (indicated by $Q$). The rate of heat released in time by the BH can be associated with the thermal part of the power radiated, namely
\begin{equation}\label{heatexchanged}
    dQ=-F_{\textrm{th}}dt\,.
\end{equation}
Moreover, in equilibrium thermodynamics context, the thermal part of the spectrum is associated with the BH temperature through the (1+1)-dimensional Stefan-Boltzmann law:
\begin{equation}\label{SBlaw}
    F_{\textrm{th}}=\frac{\pi}{12}T^2\,.
\end{equation}
However, the Stefan-Boltzmann law, Eq. \eqref{SBlaw}, and the temperature holds in the \emph{thermodynamics of equilibrium only}, leaving unclear how to define the concept of temperature, i.e., of entropy when those quantities  depend upon time.

To overcome this issue, we follow the standard procedure of defining a thermodynamic quasi-static approximation, imposing that, for  short time interval, the equilibrium is realized only \emph{locally}. Obviously, the latter cannot be realized after the critical time, $t_c$, where the mass suddenly drops.

Hence, we restrict in the interval of times given by $0<t<t_c$, where the second time derivatives of the mass can be neglected, see Fig.~\ref{ComparisonCurves}, and the flux can be approximated by
\begin{equation}\label{simplifiedflux}
    F=\frac{1}{768\pi M^2}\left(1-2\frac{\Dot{M}}{M}t+\frac{\Dot{M}^2}{M^2}t^2\right).
\end{equation}
From Eq.~\eqref{simplifiedflux}, by using Eqs.~\eqref{ans1} and \eqref{ans2}, we have for the non-thermal and thermal counterparts respectively:
\begin{subequations}
    \begin{align}
    F_{\textrm{no-th}}&=\frac{\alpha}{768\pi M^2}\left(-2\frac{\Dot{M}}{M}t+\frac{\Dot{M}^2}{M^2}t^2\right)\,,\label{nonth1}\\
    F_{\textrm{th}}&=\frac{1}{768\pi M^2}+\frac{1-\alpha}{768\pi M^2}\left(-2\frac{\Dot{M}}{M}t+\frac{\Dot{M}^2}{M^2}t^2\right).\label{th1}
    \end{align}
\end{subequations}

So, in this range of times, the temperature defined from the Stefan-Boltzmann law, Eq. \eqref{SBlaw}, is
\begin{equation}\label{temperaturebeforetc}
    T=\frac{1}{8\pi M}\sqrt{1+(1-\alpha)\left(-2\frac{\Dot{M}}{M}t+\frac{\Dot{M}^2}{M^2}t^2\right)}.
\end{equation}

\subsection{Effective temperature}

To study non-equilibrium thermodynamics in a quasi-static regime, we may define an effective temperature, valid for quite short time intervals, by
\begin{equation}\label{eq: efftemp}
    T_{\textrm{eff}}(t)=\frac{1}{\Delta t}\int_t^{t+\Delta t}T(t')dt'\,,
\end{equation}
where $T$ is the temperature defined from the Stefan-Boltzmann law, Eq. \eqref{temperaturebeforetc}.

\begin{figure}
    \centering
    \includegraphics[scale=0.7]{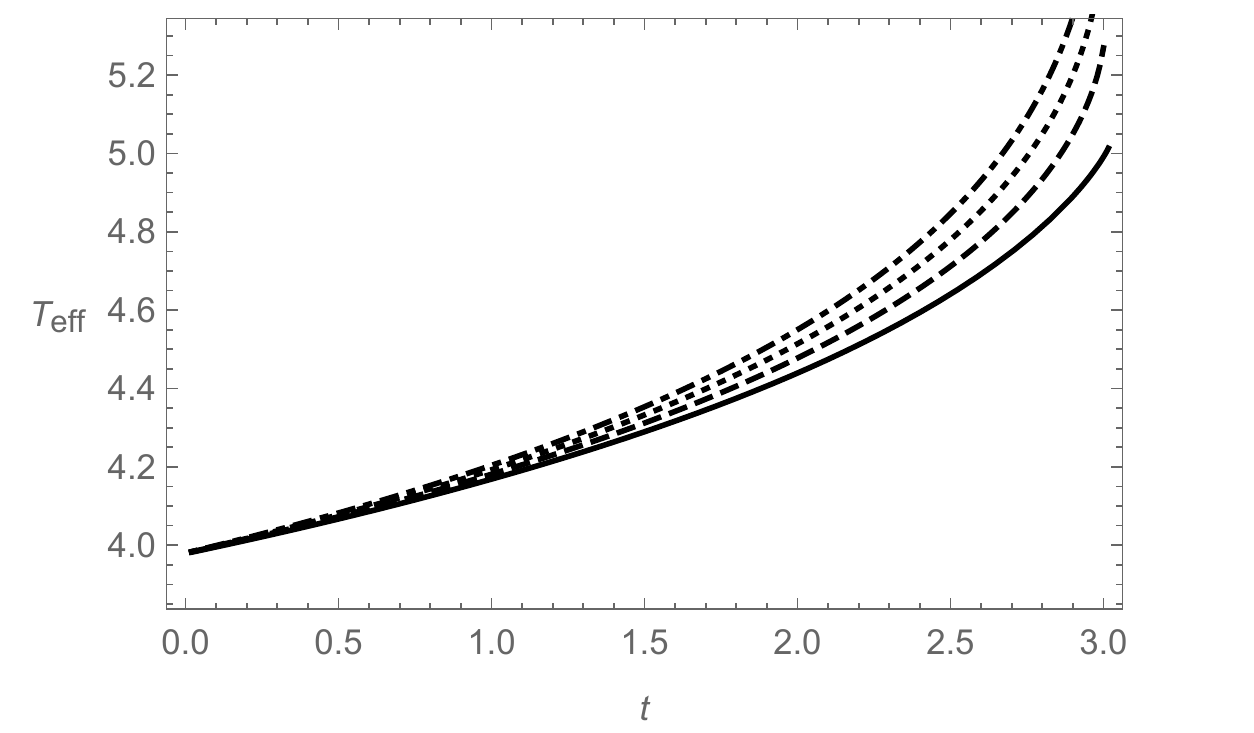}
    \caption{Plot of the effective temperature of the BH in function of time (by Eq.~\eqref{eq: efftemp}) for different values of $\alpha$. The numbers in the axes $x$ and $y$ are in powers of $10^5$ and $10^{-3}$, respectively. In particular $\alpha=1$ (thick line), $\alpha=0.95$ (dashed line) $\alpha=0.9$ (dotted line) and $\alpha=0.85$ (dot-dashed line). The integration time $\Delta t$ was chosen as $\Delta t=1000$.}
    \label{fig: effective temperature}
\end{figure}

A plot of $T_{\textrm{eff}}$ is provided in Fig.~\ref{fig: effective temperature} for different values of $\alpha$, close to unity to guarantee the quasi-static regime.
To check whether our quasi-static approximation is suitable, we compare the effective temperature, $T_{\textrm{eff}}$,  with the equilibrium temperature, namely the Bekenstein temperature \cite{Hawking:1974sw,Beck}  given by $T_{\textrm{HW}}=\frac{1}{8\pi M}$, as prompted in Fig.~\ref{fig: effective temperature}.

To certify the goodness of our hypothesis toward the quasi-static approximation, the effective temperature can be easily recast by
\begin{equation}
    T_{\textrm{eff}}=T_{\textrm{HW}}+\delta T\,,
\end{equation}
reproducing it in terms of a small deviation, globally vanishing, of the Hawking temperature, that is slightly significant for small intervals of time.
\begin{table}[h!]
\centering
\begin{tabular}{ |c|c|c|c|c| }
\hline
\hline
 $\alpha$ & $t=0.25\, t_c$ & $t=0.5\, t_c$ & $t=0.75\, t_c$ & $t=0.95\, t_c$ \\
 \hline
 \hline
 $0.95$ & $0.1423\%$ & $0.4699\%$ & $1.023\%$ & $2.820\%$\\
 \hline
 $0.90$ & $0.3286\%$ & $0.9559\%$ & $2.096\%$ & $5.567\%$\\
 \hline
 $0.85$ & $0.5146\%$ & $1.440\%$ & $3.158\%$ & $8.245\%$\\
 \hline
 \hline
\end{tabular}\caption{Values of $\delta T/T_{\textrm{HW}}$, in percentage, for specific values of $\alpha$ (indicated in the first column) and for different times $t$ (indicated in the first row).}\label{table}
\end{table}
A numerical study of $\delta T/T_{\textrm{HW}}$ is performed in Tab. \ref{table} for different times and for different $\alpha$. Since with $T_{\textrm{eff}}$ we want to approximate a thermodynamic equilibrium situation, $T_{\textrm{eff}}$ should be close to the equilibrium temperature $T_{\textrm{HW}}$, i.e., $\delta T\ll T_{\textrm{HW}}$. As we can see from Fig.~\ref{fig: effective temperature} and Tab. \eqref{table}, this occurs when $\alpha$ is close to $1$, as anticipated above.

Another relevant fact, evident from  Fig.~\ref{fig: effective temperature} and Tab. \ref{table}, is that $\delta T$ increases as $t$ approaches $t_c$. From Tab. \ref{table}, in particular, we can notice that the quasi-static approximation turns out to be still acceptable at $t=0.95 t_c$, as long as $\alpha$ is close to $1$. After $t_c$, giving the sudden increasing of the power radiated, the quasi-static approximation breaks down.

\subsection{Consequences on thermodynamics}

Once defined a quasi-static temperature, we rewrite the first law of  thermodynamics through the following assumption
\begin{equation}
dE=T_{\textrm{eff}}dS-d\Theta\,,
\end{equation}
that resembles the usual version of first thermodynamics principle, but with  $T_{\textrm{eff}}$ that replaces the equilibrium temperature, as given by Eq.~\eqref{eq: efftemp} with a corresponding net entropy\footnote{For the sake of clearness, one would require to add a subscript `$\textrm{eff}$' to the entropy also. However, we leave $S$ without any subscripts in order to  simplify the notation.}, say $S$.

Using Eq.~\eqref{heatexchanged} and $dQ=T_{\textrm{eff}}dS$ we obtain an expression for the rate of entropy loss of the BH as it evaporates:
\begin{equation}
    \frac{dS}{dt}=-\frac{F_{\textrm{th}}}{T_{\textrm{eff}}}.
\end{equation}

Integrating the last over a period of time, we obtain the entropy that the BH loses during this period. In particular, since the quasi-static thermodynamics approach is not possible after $t_c$, we study the entropy released by the BH from its creation $t=0$ (corresponding to the mass $M_0$) to $t_c$ (corresponding to the mass $M_c$), namely
\begin{equation}
    S_{\textrm{rel}}(M_0:M_c)=-\int_0^{t_c}\frac{P_{\textrm{th}}}{T_{\textrm{eff}}}.
\end{equation}
To compare the latter with the Bekenstein-Hawking entropy $4\pi M_0^2$, it is useful to write
\begin{equation}\label{eq: entropysimplified}
    S_{\textrm{rel}}(M_0:M_c)=\beta\pi M_0^2.
\end{equation}
Taking various values for $\alpha$, the corresponding findings for $\beta$, obtained from Eq.~\eqref{eq: entropysimplified}, are shown in Tab. \ref{table2}.
\begin{table}[h!]
\centering
\begin{tabular}{ |c|c|c| }
 \hline
 \hline
 $\alpha$ & $\beta$ \\
 \hline
 \hline
 $0.95$ & $1.098$\\
 \hline
 $0.90$ & $1.107$\\
 \hline
 $0.85$ & $1.112$\\
 \hline
 \hline
\end{tabular}\caption{Values of $\beta$ (indicating the entropy released during the evaporation from $M_0$ to $M_c$ from Eq.~\eqref{eq: entropysimplified}),  for different values of $\alpha$, indicating the non-thermality of the spectrum, by Eq.~\eqref{th1}.}\label{table2}
\end{table}
As we can see from this table, the more the spectrum is non-thermal (the more is $\alpha$), the less is the entropy lost by the BH during the period $0<t<t_c$ (the less is $\beta$).

We can make the reasonable assumption that the entropy of the particles radiated by the BH is proportional through a constant, say $\gamma$, to the one lost by the BH itself, i.e.,  $\frac{d S_{\textrm{rad}}}{dt}=-\gamma\frac{d S_{\textrm{rel}}}{dt}$ (see e.g.  \cite{Page:1976df,Page2emission,Page:1977um,Page_2005,Page_2013} for more information about the value of $\gamma$). In this case, from Tab. \ref{table2}, we conclude that the more non-thermal the spectrum the less entropic the BH radiation.

This result seems to be consistent with the fact that part of the information swallowed by the BH is retrievable in the eventual non-thermal part of the radiation, slightly suggesting some resolution of BH information loss.

\subsection{Consequences on entropy}

Lastly, we compare the entropies we have computed in Tab. \ref{table2} $S_{\textrm{rel}}(M_0:M_c)=\beta\pi M_0^2$ with the one released from an evaporating BH following the evaporation predicted by Hawking \eqref{M(t)Hawking}, i.e., without considering evaporation effects on the radiation, until the mass of the BH reaches $M_c$. The entropy of such a BH is given by the Bekenstein-Hawking entropy \cite{Beck}, $S=4\pi M^2$. Using the indicative value of $M_c$ provided in Eq.~\eqref{empcritvalues}, we obtain
\begin{equation}\label{BHentropy}
   \beta= \frac{S_{\textrm{rel}}(M_0:M_c)}{\pi M_0^2}\sim1.48\,.
\end{equation}
This means that, by considering the same mass evaporated $M_0-M_c$, i.e., the same amount of radiation, the Hawking radiation is more entropic than our findings in  Sec.~\ref{sec2}.

By looking at the expressions \eqref{th1} and \eqref{nonth1} for the thermal and non-thermal parts of the power radiated, respectively, we can explain qualitatively what is the further information retrievable from a BH in our case, with respect to the Hawking case. To do so, we summarize below our steps.
\begin{itemize}
    \item[-] First, as we stressed, radiation is not fully-thermal. By considering, for instance, the photon evaporation \cite{Hawking:1974sw,Page:1976df,Page_2013}, we expect that the radiated photons are no longer completely unpolarized. So, part of the information swallowed by the BH could be encoded in the polarization of the radiated photons. We confirm this fact by looking at Tab. \ref{table2}, in which we show that, the more non-thermal the radiation, the less entropy radiated as stated above. Consequently, the more the photons are polarized.
    \item[-] Suppose that we retrieve the radiated BH energy within a finite period of time while knowing its mass $M(t)$. In Hawking's case, the observed power radiated is $(768\pi M^2)^{-1}$. This contains information only about the mass $M(t)$, which we are observing. Hence, as expected, in the Hawking case we do not retrieve further information by observing Hawking radiation. Instead, by considering our model, the power radiated, in its explicit form, using Eq.~\eqref{diffeqapprox},
    is given by
    \begin{equation}
        F=\frac{384\pi M^4}{t^2}-\frac{M}{t}-\frac{384\pi M^4}{t^2}\sqrt{1-\frac{t}{192\pi M^3}}\,.
    \end{equation}
    From this expression, by observing the mass of the BH and its energy radiated, we are able to retrieve the parameter $t$, giving the time passed since the BH started to evaporate. As a consequence, while without evaporation effects on the radiation only the mass of the BH is retrievable from Hawking radiation, by considering these effects, we are able to retrieve the history of the BH mass i.e. $M(t)$. The latter implies also the information about $M(t=0)$, i.e. of the mass of the BH once it was created, and on $t$, the black hole age. This gives a decrease of the degrees of freedom of the microstates composing the BH mass, reducing the entropy as confirmed by comparing the values in Table \ref{table2} and Eq.~\eqref{BHentropy}.
\end{itemize}

\section{Outlooks}\label{sez5}

We have studied the analogy between moving mirrors and BHs, with particular attention devoted to the mass evolution of an evaporating BH in vacuum and to the corresponding non-equilibrium thermodynamics.

In particular, we adopted mirror analogs to BHs since these objects provide simplified descriptions of the time-evolving BH nature, indicating how BHs can evaporate. We described the mass evolution by means of numerical solutions obtained in the framework of Carlitz-Willey trajectory of mirrors. Consequently, we obtained suitable corrections to Hawking radiation without assuming a horizon area and/or surface gravity. We showed that these corrections are related to evaporation and we argued about possible deviation effects that appeared similar to those induced by back-reaction on the metric, investigated in previous literature. We inferred  (small) corrections to Hawking radiation, obtained as the BH evaporates, and we proposed a view of the \emph{BH information paradox} in light of our findings. Moreover, in the case of not-fully thermal radiation, we studied the non-equilibrium thermodynamics associated with BHs, passing through mirror analogs and showing, again, how to relate these outcomes to the information paradox. To do so, we worked out the hypothesis of quasi-static processes, leading to an approximate version of the first principle of thermodynamics. Deviations from Hawking radiation were computed, showing at the same time a decrease of a BH's lifetime by a factor $\sim3/8$. The entropy decrease was interpreted by assuming that part of information can be retrieved by BH radiation. Consequences about the role of an effective temperature, in view of revising the first principle of thermodynamics, has been discussed critically.

For future perspective, several aspects related to our work could be developed. One example is the role of Bogoliubov transformations in the context of mirrors, while another is the role of thermodynamics. Further applications of accelerating mirrors as BH analogs with different trajectory classes could also be pursued.

\begin{acknowledgments}
OL expresses is grateful to the Instituto de Ciencias Nucleares of the UNAM University for hospitality during the period in which this manuscript has been written. OL acknowledges the Ministry of Education and Science of the Republic of Kazakhstan, Grant: IRN AP08052311. Also, funding comes in part from the FY2021-SGP-1-STMM Faculty Development Competitive Research Grant No. 021220FD3951 at Nazarbayev University.
\end{acknowledgments}

\end{document}